\documentclass[twocolumn]{aastex631}

\newcommand{\ergs}{erg\,s$^{-1}$}
\usepackage{scalefnt}

\usepackage{multirow}
\usepackage{url}

\begin{document}

\title{A Multiwavelength View of SPLUS J102139.31-282247.5:\\ an X-ray Luminous and Eclipsing Dwarf Nova}

\author[0009-0001-7276-3571]{Amanda Silva de Araujo}
\affiliation{Departamento de Física, Universidade Federal de Sergipe, Av. Marechal Rondon, S/N, 49100-000 São Cristóvão - SE, Brazil}
\email{amandadearaujo@on.br}

\author[0000-0002-6211-7226]{Raimundo Lopes de Oliveira}
\affiliation{Departamento de Física, Universidade Federal de Sergipe, Av. Marechal Rondon, S/N, 49100-000 São Cristóvão - SE, Brazil}
\affiliation{Observatório Nacional, Rua Gal. José Cristino 77, 20921-400, Rio de Janeiro, RJ, Brazil}
\email{raimundo.lopes@academico.ufs.br}

\author[0000-0002-8286-8094]{Koji Mukai}
\affiliation{CRESST and X-ray Astrophysics Laboratory, NASA/GSFC, Greenbelt, MD 20771, USA}
\affiliation{Center for Space Science and Technology, University of Maryland, Baltimore County, 1000 Hilltop Circle, Baltimore, MD 21250, USA}
\email{koji.mukai@nasa.gov}

\author[0000-0001-5755-7000]{Raymundo Baptista}
\affiliation{Departamento de F\'{\i}sica,
Universidade Federal de Santa Catarina,
Campus Trindade,
82912-600 Florian\'opolis - SC, Brazil}
\email{raymundo.baptista@ufsc.br}

\author{Albert Bruch} 
\affiliation{Laborat\'orio Nacional de Astrof\'{\i}sica, Rua Estados Unidos, 154, 37504-364 Itajub\'a - MG, Brazil}
\email{albert@lna.br}

\author{Romualdo Eleutério}
\affiliation{Observatório Nacional, Rua Gal. José Cristino 77, 20921-400, Rio de Janeiro, RJ, Brazil}
\email{romualdojunior@on.br}

\author[0000-0003-4630-1311]{Angela Krabbe}
\affiliation{Departamento de Astronomia, Instituto de Astronomia, Geofísica e Ciências Atmosféricas da USP, Cidade Universitária, 05508-900 São Paulo, SP, Brazil}
\email{angela.krabbe@usp.br}

\author[0000-0002-5267-9065]{Claudia Mendes de Oliveira}
\affiliation{Departamento de Astronomia, Instituto de Astronomia, Geofísica e Ciências Atmosféricas da USP, Cidade Universitária, 05508-900 São Paulo, SP, Brazil}
\email{claudia.oliveira@iag.usp.br}

\begin{abstract}
We report the independent discovery of the cataclysmic variable (CV) SPLUS J102139.31-282247.5 using Southern Photometric Local Universe Survey 12 band photometry, following its initial identification via ASAS-SN variability. Here, we investigate the system using optical photometry alongside optical and X-ray spectroscopy. High-cadence optical light curves from publicly available TESS data and a dedicated campaign at the Pico dos Dias Observatory in Brazil revealed a highly inclined system exhibiting pronounced eclipses with an orbital period of 1.81 hours. We found no evidence of a WD spin period. 
Despite seasonal gaps, the long-term ASAS-SN light curve revealed several outbursts (15 with $\Delta$mag $\sim$ 1.5 brighter than the mean magnitude) and one superoutburst over the past decade. During Swift X-ray telescope observations, the system exhibited X-ray luminosity of $8.5 \times 10^{31}$ erg s$^{-1}$ in the 0.3-10\,keV band, originating from thermal plasma emission at a temperature consistent with accretion shocks. Optical spectroscopy from the Gemini Multi-Object Spectrographs on the Gemini South telescope confirms the CV nature of the system, showing strong, double-peaked H and He\,I emission lines alongside a weak He\,II\,4686 line. 
Altogether, these features strongly support the classification of SPLUS J102139.31-282247.5 as a quiescent, eclipsing dwarf nova with an unusually high X-ray luminosity.

\end{abstract}

\keywords{Close binary stars (254) --- Eclipsing binary stars (444) ---
Cataclysmic variable stars (203)}

\section{Introduction} 
Whether isolated or in binary systems, white dwarfs (WDs) are key laboratories for stellar astrophysics \citep[see][]{anabranca,2022Galax..10...43S}. Among those in binary systems are the ones that belong to interacting systems, in an evolutionary phase in which they accrete matter from a low-mass, main-sequence companion that overflows its Roche lobe: the Cataclysmic Variables \citep[CVs; see][for a review]{mukai2017}. 
This configuration adds new significance to these compact objects (10$^{3-6}$\,g\,cm$^{-3}$), which span a wide range of magnetic field strengths (from 1 kG to 1000 MG). Combined with the WD rotation and the amount of matter accreting onto it, the magnetic field intensity sets a rich environment in phenomenology following the different accretion modes. 
Those hosting a weakly magnetic ($\leq 100$\,kG) WD allow the formation of an accretion disk that almost reaches the stellar surface while driving outbursts \citep[like in dwarf nova systems;][]{warner}. At the extreme are the strongly magnetized WDs, forming ``polar'' systems. They are known as AM Herculis systems, in which the ionized material follows the magnetic field lines without formation of an accretion disk and then reaches the conditions of high density and high speed that define a magnetized accretion column(s) \citep{2025arXiv250510337S}.\footnote{https://www.aip.de/de/members/axel-schwope/polarcat} It produces polarized and cyclotron radiation, as well as bremsstrahlung radiation that includes the occurrence of shock followed by cooling flow as it approaches the WD's surface \citep{cropper, patterson}. At intermediate magnetic field intensities, both components, the accretion disc and the accreting column(s), are present, as is the case for ``intermediate polar'' (IP) systems.\footnote{https://asd.gsfc.nasa.gov/Koji.Mukai/iphome/catalog/alpha.html} 

The connection of the subclasses and how their systems evolve is far from being a settled matter. The discussions include the nature and how the WD magnetic field changes over time \citep[e.g.,][]{2021NatAs...5..648S}, and details on the way that CVs shift from longer ($\sim$\,15\,hrs) to shorter ($\sim$\,1\,hr) orbital periods \citep[e.g.,][]{2022MNRAS.513.4169S}, finally bouncing to longer periods again \citep[e.g.,][]{2023A&A...679L...8S}. 
As for the big picture, eruptions from the WD surface chemically enrich and inject energy to the interstellar medium \citep[e.g.,][]{starrfield2017}. Additionally, accreting white dwarfs (AWDs) may explode as Type Ia supernovae, thereby establishing a direct connection between stellar astrophysics and cosmological domains \citep[e.g.,][]{maoz2014}.

A major challenge in AWDs lies in the underrepresentation of systems across various subclasses, resulting in limited coverage of the parameter space where physical properties are found. There is significant divergence between observations and theoretical predictions \citep[e.g.,][]{Pretorius}. Part of such difficulty is due to observational biases.

Recent advances in observational astronomy, as through wide-field surveys like the All Sky Automated Survey for SuperNovae \citep[ASAS-SN;][]{Kochanek_2017,shappee}, 
Catalina Real-Time Transient Survey \citep[CRTS;][]{Drake_2009,10.1093/mnras/stu639},
Optical Gravitational Lensing Experiment  \citep[OGLE; e.g.,][]{2013AcA....63..135M}, 
Transiting Exoplanet Survey Satellite \citep[TESS;][]{Ricker_2014}, GAIA Survey \citep{2016A&A...595A...1G}, and extended ROentgen Survey with an Imaging Telescope Array \citep[eROSITA;][]{erosita}, have significantly contributed to increasing the number of known AWDs. Recent efforts have emerged in the field, like the Southern Photometric Local Universe Survey offering 12-band photometry \citep[S-PLUS;][]{LopesdeOliveira_2025} and the expectation of the Vera C. Rubin Observatory \citep{2019ApJ...873..111I}.
 The findings go beyond the increase in the number of systems, but the revelation of exotic cases constrains theoretical models and offers deeper insights into the physical mechanisms that govern mass transfer and accretion in binary systems. 

We report an independent discovery of an accreting WD system, SPLUS J102139.31-282247.5 (hereafter referred to as SPLUS J102139) -- after the citation in the International Variable Star Index \citep[VSX;][]{vsx}, as a CV from photometric variability detected with the ASAS-SN (ASASSN-18cv). The system was pinpointed during the pilot project with the S-PLUS reported by \cite{LopesdeOliveira_2025}. 
Here we characterize the system through optical photometry and spectroscopy, and X-ray spectroscopy, from S-PLUS, Gemini South, Pico dos Dias Observatory (OPD), and The {\it Neil Gehrels} Swift Observatory \citep[Swift;][]{swift}, added to public information from ASAS-SN, TESS, and GAIA.

\section{Observations and Data Reduction}

\subsection{Multiband Photometry: The Southern Photometric Local Universe Survey (S-PLUS)}
\label{sct:obssplus}

S-PLUS is a project aimed at mapping 9,300 square degrees of the sky in twelve photometric bands from $\sim$\,3500\,\AA\ to $\sim$\,10000\,\AA: five broad ($ugriz$) and seven narrow-band filters  (J0378, J0395, J0410, J0430, J0515, J0660, and J0861). The broad-band filters were optimized to sample key spectral features such as the Balmer jump, H$\delta$, H$\alpha$, and the calcium triplet \citep{splus}.

\begin{figure}[t!]
\hspace*{-0.1cm}\includegraphics[width=8.4cm,scale=0.1]{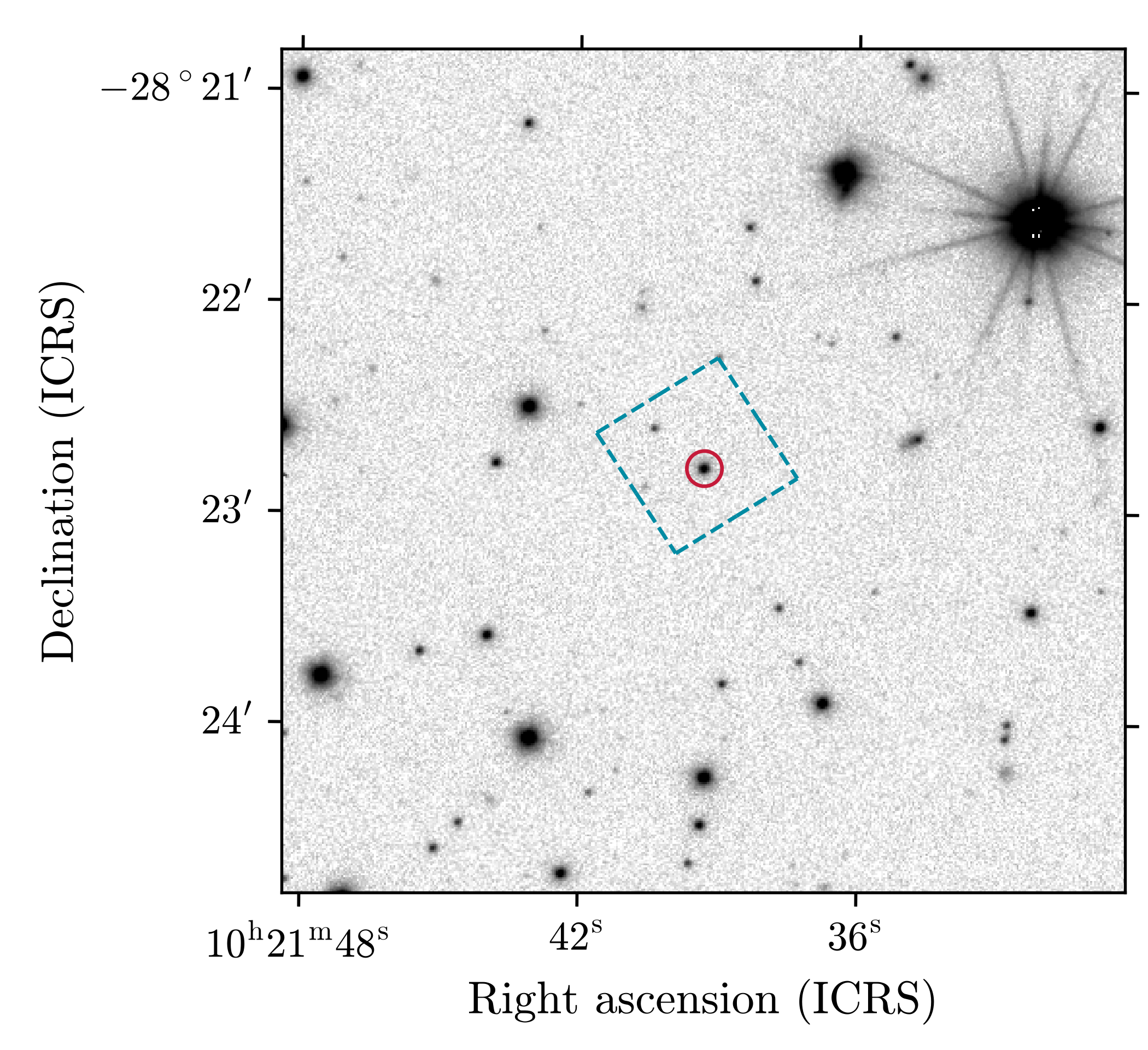}
\caption{S-PLUS $r$-band image covering an approximately $4 \times 4\text{ arcmin}$ region around SPLUS J102139. The target position is indicated by the red circular region (with an arbitrary radius), while the cyan rectangle marks the aperture used to extract its Sector 36 TESS light curve.}
\label{fig:splusimage}
\end{figure}

S-PLUS offers an unprecedented collection of colours and spectral energy distributions (SEDs) for targets in the Southern celestial hemisphere. Those from the third S-PLUS data release ($\sim$\,1800 deg$^2$) were explored by \citet{LopesdeOliveira_2025} in the context of identifying accreting WDs, and  SPLUS J102139 was one of them -- the 11th system, highlighted as “deserving a separate paper with a more in-depth investigation.” The identification of accreting WDs {\bf by} \citet{LopesdeOliveira_2025} was based on two main criteria: evidence of H$\alpha$ emission\footnote{[(mag$_r$+mag$i$)/2 - mag${J0660}$] $>$ 0.1 mag; note the typo in the equation presented in \citet{LopesdeOliveira_2025}} and a “V-shaped” SED, suggestive of a composite system formed by a hot accretor and a cooler donor star. The {\bf SEDs showed} a dip separating a blue excess ($<$\,5000\AA), attributed to the WD's surface and the material under accretion, from a red excess ($>$\,5000\AA), associated with a cool companion and possibly an accretion disk. After {\bf their} identification with the S-PLUS photometry, we initiated spectroscopic follow-up observations with Gemini and photometric observations using OPD resources, while also analyzing public data archives. Altogether, these efforts form the foundation of this study {\bf on SPLUS J102139 (see Fig. \ref{fig:splusimage})}.

\subsection{Gaia Mission}
\label{sct:gaia}

To accurately determine the distance and astrometric properties of SPLUS J102139, we utilized data from the Gaia mission \citep{2016A&A...595A...1G}, specifically from its Data Release 3 (DR3). The angular distance between the exact S-PLUS coordinates and Gaia coordinates is extremely small, measuring approximately 0.10''. The Gaia astrometric solution yields a parallax of $1.99$ mas and proper motions of $\mu_\alpha \cos \delta = -19.38$ mas yr$^{-1}$ and $\mu_\delta = 0.26$ mas yr$^{-1}$. We adopted the geometric distance of $499_{-19}^{+23}$ pc derived from Gaia parallax measurements as reported by \citet{gaiadr3}. This reliable distance determination was crucial to estimate the X-ray luminosity of the system observed with the Swift telescope (see \ref{sect:xrt_results}).

Finally, the mean apparent Gaia magnitude in the $G$ band is $m_G = 17.74$. To estimate the absolute $G$ magnitude ($M_G$), we subtract from the apparent magnitude the corresponding distance modulus and the $G$-band extinction, $A_G$. Using $E(B-V) = 0.034$ estimated from 3D extinction maps \citep{lallement}, and adopting $A_G/A_V \approx 0.90$ with $A_V \approx 3.1 \times E(B-V) \approx 0.105$, we find a very small extinction of $A_G \approx 0.095$. This results in an absolute magnitude of $M_G \approx 9.16$, which is consistent with the expected brightness of a system in quiescence.

\subsection{Optical Spectroscopy: Gemini South}
\label{sct:obsgemini}

Optical spectroscopy of SPLUS J102139 was obtained with the Gemini South Multi-Object Spectrographs (GMOS-S) on April 15, 2023 (GS-2023A-Q-411 Poor Weather program; PI: R. Lopes de Oliveira). The observations were performed with a single long-slit with a width of 1.0'', resulting in a spectral resolution of approximately 5.6 \AA. The B600-G5323 grating (ruling density of 600 lines per mm) was employed at two central wavelengths ($\lambda_c$) in four exposures of 600 seconds each: two having $\lambda_c$\,=\,510 nm and the other two with $\lambda_c$\,=\,570 nm. This complementary dual setup resulted in a continuous spectral coverage spanning from $\sim$\,3560 \AA\ to $\sim$\,7300 \AA\ -- their distinct central wavelengths compensates for gaps between CCDs and their defective columns.

Data reduction and the combination of spectra from the applied grating configurations were performed using the DRAGONS software (v3.1.0; \citealt{dragons}), which is specifically developed for Gemini data. 
The {\fontfamily{qcr}\selectfont reduce} task was used to apply bias subtraction, flat-field correction, bad-pixel masking, cosmic ray removal, and wavelength calibration using arc lamp exposures. Flux calibration was carried out using observations of the standard star LTT7987. 
The target was observed under poor weather conditions and the reference star was observed at a different epoch (as part of Gemini’s weekly calibration data). Therefore, due to varying weather and airmass conditions, we obtained relative rather than absolute fluxes -- which are sufficient for our purposes. The optical spectrum is presented in Fig. \ref{fig:gemini}.

\subsection{Optical light curves} 

Amplitude variations in minute- and day-time-scale brightness of SPLUS J102139 were assessed using optical light curves from a campaign conducted at the OPD, and from the TESS and ASAS-SN publicly available databases. The dataset explored in this work is described in detail in the following subsections.

\subsubsection{The All-Sky Automated Survey for Supernovae (ASAS-SN)}

The ASAS-SN project currently operates 24 telescopes distributed across both the Northern and Southern Earth hemispheres, enabling regular monitoring of the entire sky. While its primary goal is to detect bright supernovae, the survey has also proven effective in uncovering and monitoring a variety of other transient and variable sources. This was the case of ASASSN-18cv, referred to here as SPLUS J102139. This system has a publicly available, daily-cadence ASAS-SN light curves in the visual range (V- and g-bands) spanning from November 22, 2013, to November 25, 2024 (see Fig. \ref{fig:asas}). About one-third of the available photometry is limited to upper limits or has undetermined uncertainties and was excluded from the analysis. With the remaining data, even considering seasonal gaps and data exclusions, it is still possible to assess the source's behavior over 11 years. The uncertainties in the photometry are below 0.22 magnitudes. As discussed in Section \ref{sct:asassnlc}, the analysis allowed for the identification of outbursts and superoutbursts.

\subsubsection{The Transiting Exoplanet Survey Satellite (TESS)}

The TESS mission provides photometry of wide-field regions of the sky (24$\times$96 deg$^{2}$) per exposure, each one defining one ``sector" on the celestial sphere and typically being observed over two satellite orbits ($\sim$\,2$\times$13.7 days). Except for periods when the satellite is not observing --  typically during data downlinks to Earth or while in safe mode -- it provides uninterrupted light curves. 

TESS observed SPLUS J102139 in four sectors: 36, 62, 63, and 89. Sector 36 was observed for 25.09 days in 2021, from March 7 to April 1, when the source was located on CCD4 of Camera 2. In 2023, observations of the source during Sectors 62 (CCD3 of Camera 2) and 63 (CCD4 of Camera 2) provided an additional 52.24 days of coverage, from February 12 to March 10 and from March 10 to April 6, respectively. Sector 89 was observed from February 11 to March 22, 2025, with the source positioned on CCD3 of Camera 2, contributing an additional 28.12 days of data. Data gaps occurred approximately every 14 days in Sector 36, and every about 7 days in Sectors 62, 63, and 89. All observations were conducted with a cadence of 120 seconds.

TESS provides two types of flux measurements: Single Aperture Photometry (SAP) Flux and Pre-search Data Conditioned Simple Aperture Photometry (PDCSAP) Flux. Although PDCSAP Flux corrects for instrumental variations, it can sometimes affect the light curves of systems like dwarf novae during outbursts \citep{bruch} or introduce noise that mimics real transit signals in exoplanets \citep{unknown}. Since instrumental effects do not significantly impact the frequency range $> 1/d$ investigated in this study \citep{sapflux}, we used the SAP flux for our analysis. 

 We identified systematic differences in the overall flux level between sectors. These offsets likely result from instrumental effects, such as variations in the photometric zero point or changes in background contamination across different sectors.
To account for these inter-sector offsets and ensure consistency across the full light curve, we normalized the data by dividing the raw flux by the median flux of each dataset. This ensures that flux variations are expressed relative to a consistent baseline, enhancing the visibility of timing behavior while minimizing the impact of absolute flux differences. This normalization allows us to focus on timing characteristics without interference from variations in the flux scale; it is the relative variation, not the absolute values, that is of interest to us. The normalized TESS light curves of SPLUS J102139 are shown in the top panel of Fig. \ref{fig:tess}.

\subsubsection{The Pico dos Dias Observatory}

We observed SPLUS J102139 with a 1.6-m telescope at the Pico dos Dias Observatory, in Brazil, operated by the National Laboratory for Astrophysics. The observations were obtained using the Simultaneous Polarimeter and Rapid Camera (SPARC4) in the photometric mode, which enables simultaneous fast photometry in the SDSS-like $g$, $r$, $i$, and $z$ bands \citep{2012SPIE.8446E..26R,2025PASP..137c5003B}. 

The SPARC4 observations were conducted on 1 March 2025 under intermittent cloud cover. Measurements in such a bad weather conditions and affected by technical issues were excluded from the final light curves. The exposure time for each band was independently set as a compromise between maximizing the signal-to-noise ratio and maintaining short time binning for improving temporal resolution. The exposures for the $r$ and $i$ bands were set to 10 seconds. For $g$ and $z$, the less sensitive bands, we used 30-second exposures. 

The data reduction followed the standard procedure with the PRAIA software \citep[PRAIA;][]{praia} and Image Reduction and Analysis Facility \citep[IRAF;][]{iraf1,iraf2}\footnote{https://iraf-community.github.io/}, with the classical bias and flat-field corrections. Absolute photometric calibration  was not required for the objectives of this study. Since we are interested in the relative variability of the source, meaning that the magnitude difference is what matters, we adopted differential photometry using a comparison star. The reference star was Gaia DR3 5462557561328753792, which has a G-band magnitude of 17.0. 

The SPARC4 light curves are shown in Fig. \ref{fig:sparc4}. Uncertainties in the data points across all filters are below 0.06 mag. These light curves enabled us not only to follow eclipses and flickering of the system on a timescale shorter than that of the TESS light curves, but also allowing us to check for any dependence of them in four photometric bands.

\subsection{X-ray spectroscopy and UV photometry: Swift}
The \textit{Swift} telescope observed SPLUS J102139 on April 15, 21, and 26, 2023 (PI: R. Lopes de Oliveira) with the X-ray Telescope in Photon Counting mode and with the Ultra-violet Optical Telescope (UVOT) with the ``filter of the day''.
 Table \ref{tab:swiftdata} summarizes the details of the observations, including magnitude and flux values. The X-ray observations
00015951001 and 00015951003 accumulated too few source counts and were not useful for spectral investigations.

The \textit{Swift-XRT} data were processed to produce cleaned and calibrated
event files using the {\fontfamily{qcr}\selectfont
xrtpipeline} of the HEAsoft software version 6.30.1, with calibration files CALDB version 20220331. The source image and spectrum were derived using {\fontfamily{qcr}\selectfont
xselect} task. The source contribution was determined from a circular region with a radius of 25'' centered on the target, while the background was estimated using a nearby source-free circular region with a radius of 39''. The {\fontfamily{qcr}\selectfont
xrtmkarf} task was employed to create an ancillary response file, correcting for counts lost due to hot columns and bad pixels, and the response matrix file spectrum. To ensure reliable statistical analysis and a balance between adequate signal-to-noise ratio and spectral resolution, we grouped the data to a minimum of 5 counts per bin using the task {\fontfamily{qcr}\selectfont grppha}. 
All spectral modeling and parameter estimation were carried out in {\fontfamily{qcr}\selectfont XSPEC} version 12.12.1 \citep{xspec}, using the grouped spectra and associated response files.
The {\fontfamily{qcr}\selectfont setplot rebin 2 100} command in {\fontfamily{qcr}\selectfont XSPEC} was used to rebin the spectrum for visualization purposes -- combining adjacent bins to have a significant detection of at least 2\,$\sigma$ but in sets with no more than 100 bins each.

Data reduction of \textit{Swift/UVOT} observations was also performed with the HEAsoft software. Each observation resulted in multiple image extensions, corresponding to different exposures. To enhance source detection, the contents of these extensions were summed using the {\fontfamily{qcr}\selectfont uvotimsum} tool. {\fontfamily{qcr}\selectfont
Uvotmaghist} task was applied to perform aperture photometry and derive magnitude, error, count rate, and flux values of the source in each observation (see Table \ref{tab:swiftdata}).

\section{Results}

\subsection{Optical spectral features}

\begin{figure*}[ht!]
\hspace*{-0.1cm}\includegraphics[width=17.74cm,scale=0.1]{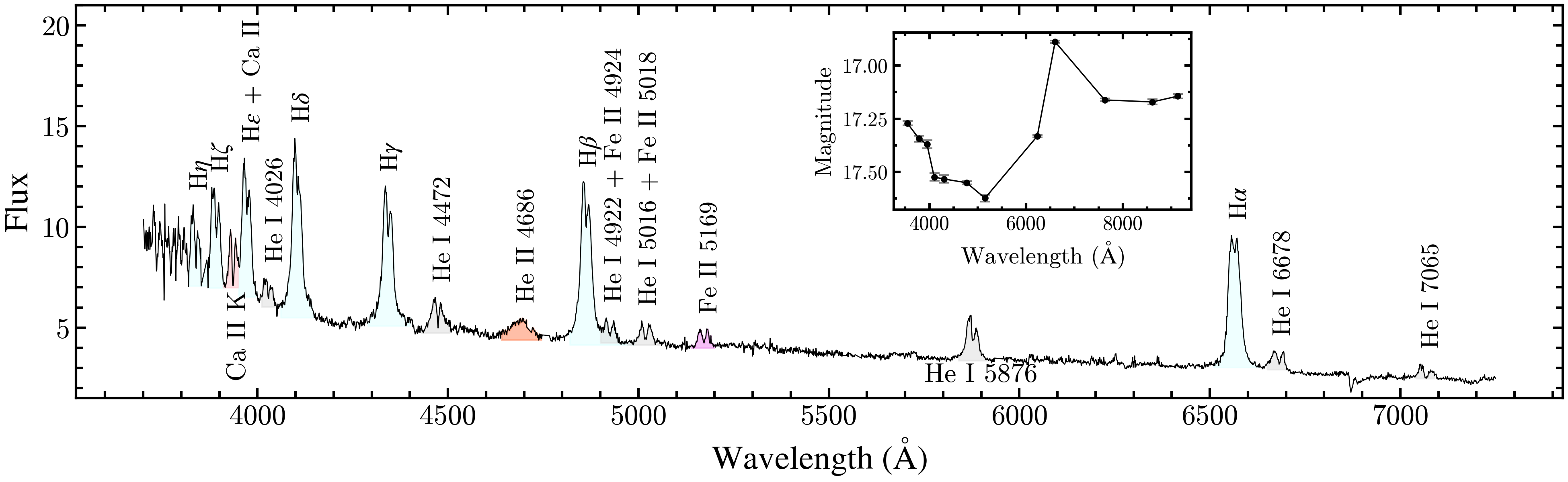}
\caption{Gemini South optical spectrum of SPLUS J102139. The inset shows the S-PLUS spectral energy distribution (SED), highlighting the photometric measurements in 12 bands. Flux is in units of $10^{-16}\,\mathrm{erg\,s^{-1}\,cm^{-2}\,\AA^{-1}}$.}
\label{fig:gemini}
\end{figure*}

\begin{table}[ht]
    \centering
    \scalefont{0.95}
    \begin{tabular}{ccc}
         \hline
         \hline
         Line & EW & FWHM \\
              & (\AA) & (\AA) \\
         \hline
         H$\eta$                       & -10.72 $\pm$ 0.53 & 21.00 $\pm$ 0.82 \\[2pt]
         H$\zeta$                      & -16.23 $\pm$ 2.96 & 21.77 $\pm$ 0.56 \\[2pt]
         Ca\,II K                      &  -6.88 $\pm$ 0.24 & 19.85 $\pm$ 0.42 \\[2pt]
         H$\epsilon$ + Ca II H                 & -24.10 $\pm$ 0.94 & 22.35 $\pm$ 0.85 \\[2pt]
         He\,I 4026                    &  -4.28 $\pm$ 0.25 & 22.25 $\pm$ 1.00 \\[2pt]
         H$\delta$                     & -39.78 $\pm$ 1.23 & 29.08 $\pm$ 0.61 \\[2pt]
         H$\gamma$                     & -45.50 $\pm$ 1.93 & 32.91 $\pm$ 0.75 \\[2pt]
         He\,I 4472                    &  -9.64 $\pm$ 0.55 & 32.04 $\pm$ 1.54 \\[2pt]
         He\,II 4686                   & -11.58 $\pm$ 0.52 & 60.68 $\pm$ 2.06 \\[2pt]
         H$\beta$                      & -60.05 $\pm$ 1.62 & 33.91 $\pm$ 0.67 \\[2pt]
         He\,I 4922 + Fe\,II 4924      &  -6.67 $\pm$ 0.19 & 26.94 $\pm$ 0.55 \\[2pt]
         He\,I 5016 + Fe\,II 5018      &  -4.15 $\pm$ 0.15 & 27.16 $\pm$ 0.58 \\[2pt]
         Fe\,II 5169                   &  -5.75 $\pm$ 0.17 & 25.33 $\pm$ 0.57 \\[2pt]
         He\,I 5876                    & -16.00 $\pm$ 0.44 & 27.48 $\pm$ 0.61 \\[2pt]
         H$\alpha$                     & -83.47 $\pm$ 2.83 & 33.89 $\pm$ 0.55 \\[2pt]
         He\,I 6678                    & -13.34 $\pm$ 0.50 & 35.55 $\pm$ 1.26 \\[2pt]
         He\,I 7065                    &  -5.84 $\pm$ 0.23 & 36.18 $\pm$ 0.71 \\[2pt]
         \hline
    \end{tabular}
    \caption{Equivalent widths and full widths at half maximum
    of the 
    optical emission lines identified in the Gemini/GMOS-S spectrum of SPLUS J102139. }
    \label{tab:linhas}
\end{table}

As mentioned in Section \ref{sct:obssplus}, SPLUS J102139 was independently identified during the pilot project presented in \citet{LopesdeOliveira_2025}. The photometric S-PLUS SED of SPLUS J102139 is shown in the inset of Fig. \ref{fig:gemini}, which also presents the optical spectrum obtained with the GMOS instrument on the Gemini South telescope (Section \ref{sct:obsgemini}). 

Optical spectroscopy confirms the expectations raised by the photometric SED, revealing the spectral features in rich detail (see Fig. \ref{fig:gemini}). 
A contribution to a rising blue flux at wavelengths below $\sim$\,5500\,\AA\ is quite noticeable. 
Notably, the absence of broad Balmer absorption troughs indicates that this blue continuum does not primarily originate from the stellar surface. Instead, it is characteristic of optically thin emission (resulting in a Balmer jump in emission) originating from the surface of the accretion disk.
This continuum is accompanied by strong emission lines, notably from Balmer series hydrogen transitions - with all lines from H$\alpha$ to H$\eta$ being visible - along with numerous He\,I and the He\,II 4686\,\AA\, Ca\,II K, and Fe\,II multiplet 42 lines. The decline in flux that would follow from the WD's light alone to longer wavelengths is interrupted. This is naturally explained by the additional contribution of a continuum emission that arises from the companion star and the accretion disk. 

The presence of an accretion disk is supported by the double-peaked emission lines. This suggests that the disk and the orbital plane are viewed at a close-to-edge-on orientation, leading to the expectations that the system is eclipsing (e.g., see sections \ref{sct:asassnlc} and \ref{sect:orbmodul}).
Using \textit{SpecUtils}, a \textit{Python} package, we measured the equivalent width (EW) and full width at half maximum (FWHM) of the strongest emission lines (Table \ref{tab:linhas}).
The high-excitation He\,II\,4686 \AA\, line is commonly used as a proxy for highly magnetized WDs in CVs, due to the characteristic plasma conditions in these systems \citep{mukai2017}. In particular, \citet{silber} proposed the following empirical criterion: magnetic CVs tends to have $|EW(H\beta)|$ exceeding 20\,\AA\, and the $EW(HeII \lambda4686)/EW(H\beta)$ ratio is greater than 0.4. For SPLUS J102139, the $EW(HeII\lambda4686)/EW(H\beta)$ ratio is about 0.2, well below this threshold. While the values do not rule out the system being an intermediate polar, they support a non-magnetic classification.

\subsection{Transient Phenomena: Outbursts, Superoutbursts and Irregular Events}
\label{sct:asassnlc}

\subsubsection{From the ASAS-SN light curve}

\begin{figure*}[ht!]
\centering
\includegraphics[width=1.01\textwidth]{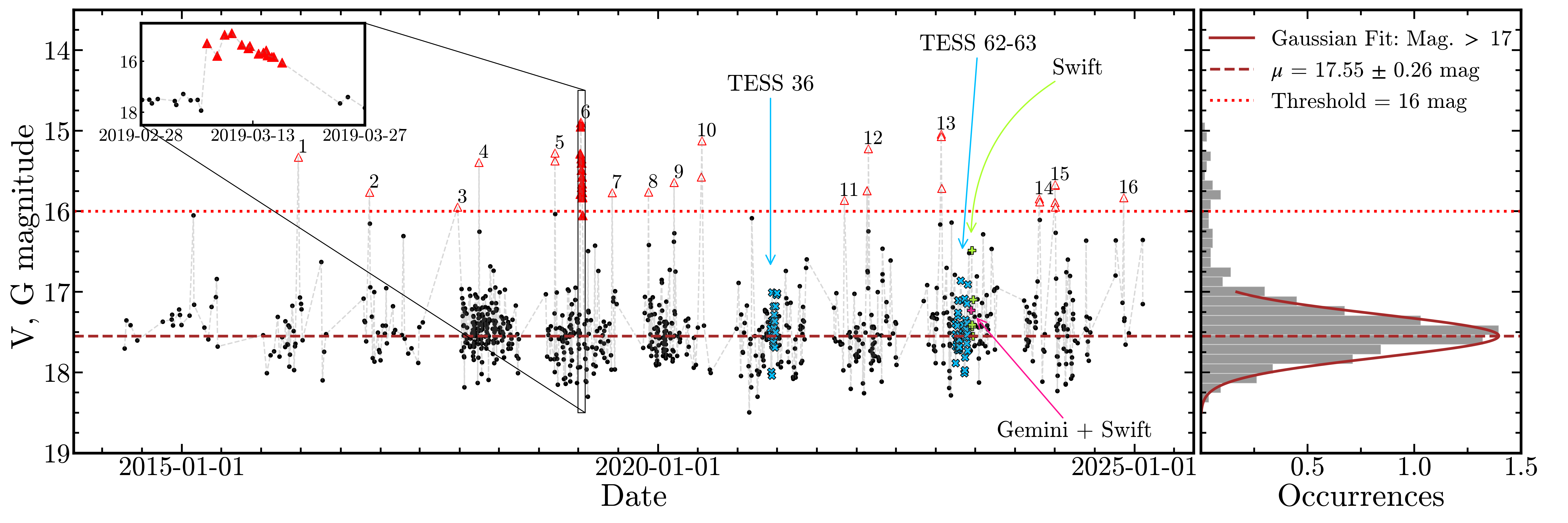}
\caption{ {\it Left: }ASAS-SN light curve in the V band. The arrows indicate the start time of the TESS (in blue), Swift (green), and Gemini  and Swift (pink) observations. Blue, green and pink crosses show ASAS-SN magnitudes within the time windows of TESS, Swift and Gemini/Swift, respectively. The open red triangles mark the identified outbursts, while the closed red triangles correspond to the superoutburst (see text). These events are labeled and listed in Table~\ref{tab:eventdata}, with the superoutburst region shown in detail for better visualization. {\it Right: }Histogram of the V band magnitudes. The solid brown line represents a Gaussian fit considering only magnitudes fainter than 17. In both panels, the brown dashed line indicates the centroid ($\mu$) of the Gaussian fit, and the red dotted line marks the 16 mag threshold adopted for outburst selection.}
\label{fig:asas}
\end{figure*}

\begin{table}[ht!]
\centering
\begin{tabular}{cccc}
\hline \hline
Event & Start Date & Duration & Peak Magnitude \\
 & (UTC) & (days) & (mag) \\
\hline
1 & 2016-03-22 & $< 8.2$ & 15.32 \\
2 & 2016-12-20 & $< 8.9$ & 15.76 \\
3 & 2017-11-23 & $< 6.0$ & 15.95 \\
4 & 2018-02-13 & $< 3.0$ & 15.39 \\
5 & 2018-12-01 & $< 4.3$ & 15.37 \\
6 & 2019-03-08 & $< 12.8$ & 14.90 \\
7 & 2019-07-08 & $< 6.0$ & 15.77 \\
8 & 2019-11-25 & $< 4.2$ & 15.76 \\
9 & 2020-03-02 & $< 3.9$ & 15.64 \\
10 & 2020-06-14 & $< 15.9$ & 15.58 \\
11 & 2021-12-14 & $< 10.2$ & 15.87 \\
12 & 2022-03-12 & $< 12.0$ & 15.74 \\
13 & 2022-12-21 & $< 24.0$ & 15.71 \\
14 & 2024-01-01 & $< 14.1$ & 15.88 \\
15 & 2024-03-01 & $< 44.0$ & 15.95 \\
16 & 2024-11-20 & $< 5.0$ & 15.83 \\
\hline
\end{tabular}
\caption{Notable outburst events observed in SPLUS J102139 during ASAS-SN observations. The durations are presented as upper limits, defined by the time elapsed between the last quiescent data point ($V > 17$ mag) before the outburst and the first one after it.}
\label{tab:eventdata}
\end{table}
Photometry of SPLUS J102139 collected over 11 years by the ASAS-SN survey has revealed significant variability in its brightness, with the magnitude ranging from 14.90 to 18.49 in the V and g bands (Fig. \ref{fig:asas}). The basal emission is dominated by variability on a one-day timescale or less, following a Gaussian distribution that peaks at 17.55 mag when only events above 17 mag are considered, with $\sigma= 0.26$\,mag
 (see right panel in Fig. \ref{fig:asas}). This value is what we define in this work as the ``quiescent level'' of the system. Episodes of sudden brightness increases punctuate that component. A total of 16 such events were recorded, occurring on average every 180 days. These events, numbered and highlighted in Fig. \ref{fig:asas} as open triangles, were identified based on a threshold of 16 mag.
Table \ref{tab:eventdata} summarizes their dates of occurrence, durations, and peak magnitudes, subject to the limitations imposed by the available temporal coverage. The durations are quoted as upper limits. These limits are defined as the time difference between the last quiescent data point ($V > 17$ mag) immediately preceding the outburst and the first quiescent data point following it. To illustrate this approach, Event 15 is reported with a duration of $< 44.0$ days. This large value does not imply the outburst itself lasted for over a month, but rather reflects a 39-day seasonal gap in the ASAS-SN coverage immediately preceding the detection of the eruption. Since the exact onset of the outburst within this unobserved window is unknown, the quoted value represents the maximum possible duration bounded by the nearest quiescent observations. Furthermore, seasonal gaps in the data acquisition and observations with low-quality or unusable data imply that some episodes may have been missed and that our estimate of the outburst recurrence interval may be an upper limit.

The most remarkable outburst took place in March 2019 (Event 6), when the system reached a V magnitude of 14.9 and remained brighter than $\sim$\,16.0 for over 12.8 days. It is highlighted in the light curve with solid red triangles, and a zoomed-in inset was added to enhance the visibility of the decay phase. This is a characteristic behavior of superoutbursts in SU UMa-type, which occur when a thermally unstable accretion disk grows large enough to trigger a tidal resonance with the companion star \citep{osaki1989}. 

Events 12 and 13, observed in March and December 2022, respectively, did not match the classic superoutburst profile in duration but stood out due to their intermediate durations (5 and 2 days, respectively) compared to the typical 1-day normal outbursts. While the sparse daily sampling of the ASAS-SN light curve prevents us from resolving their detailed profiles or confirming any complex brightness substructure during these brief events, their intermediate timescales suggest they could represent modulated or possibly aborted superoutbursts. In such scenarios, the disk may have approached near-critical conditions for tidal instability without fully entering a sustained resonant state, a behavior predicted by the thermal–tidal instability model when variations in mass transfer or incomplete disk expansion prevent the development of a full superoutburst cycle \citep{Osaki1996}. In contrast, the remaining events were very short (1-2 days) and reached more modest peak magnitudes, consistent with normal outbursts triggered solely by thermal instabilities in the accretion disk.

\subsubsection{From TESS light curves}

\begin{figure*}[t!]
\hspace*{-0.5cm}\includegraphics[width=1.04\textwidth]{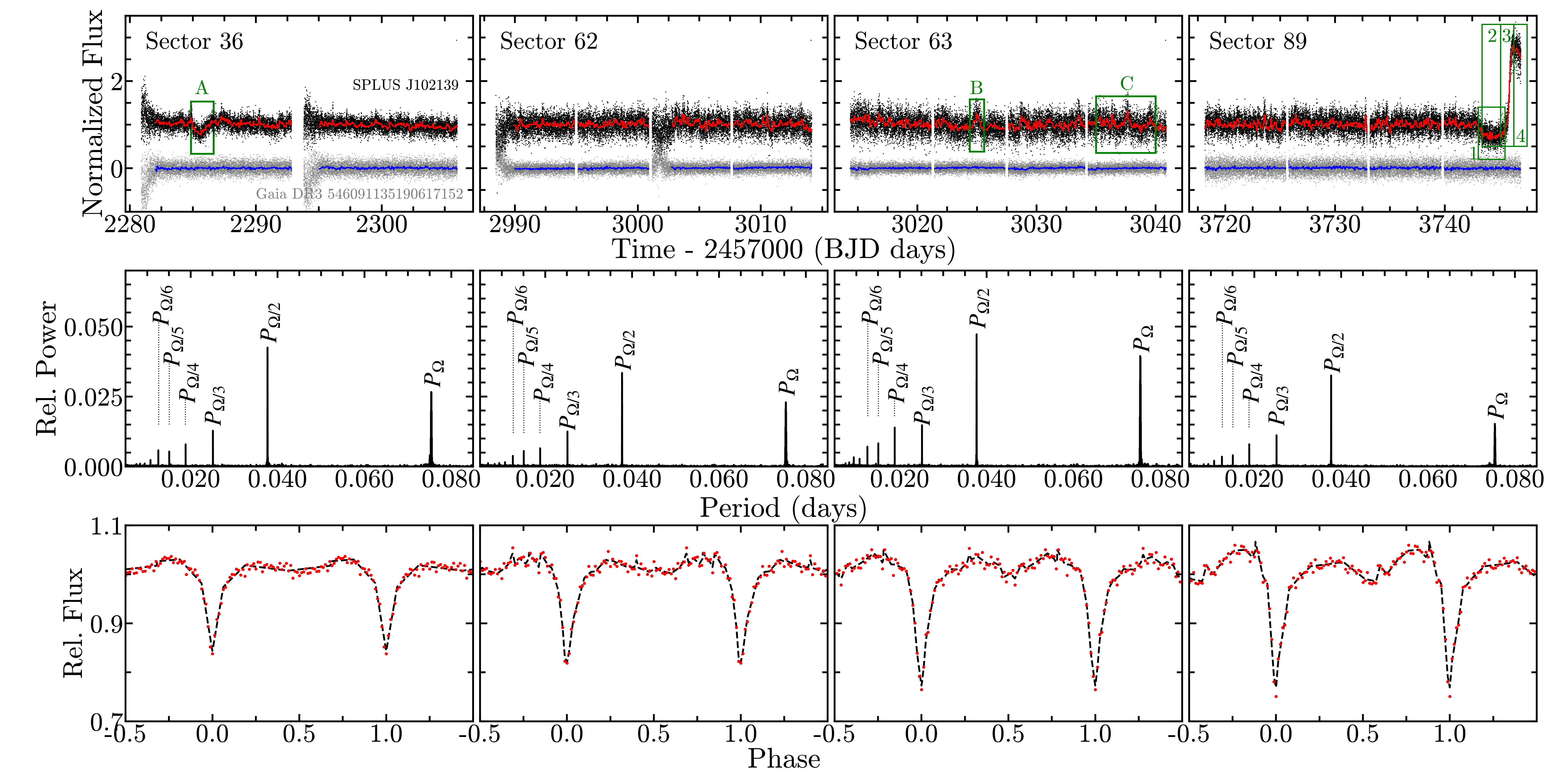}
\caption{Overview based on TESS data from observations of SPLUS J102139. 
\textit{Top:} TESS light curves of SPLUS J102139  (black) and the star Gaia DR3 5460911351906171520 (gray). The light curve of Gaia DR3 5460911351906171520 was shifted by –1 for better visualization. 
\textit{Middle:} Lomb-Scargle periodograms. 
\textit{Bottom:} TESS light curves phase-folded at 1.81 hours, with bins of 0.0125 in phase for clarity.}

\label{fig:tess}
\end{figure*}

SPLUS J102139 was observed by TESS in four sectors (36, 62, 63, and 89), with the first three at specific moments during the ASAS-SN coverage period available during the writing of this paper -- their start times are marked in Fig. \ref{fig:asas}. The top panel of Fig. \ref{fig:tess} presents the light curves with two different binnings: 2 min (the original bin) in black and 90 min in red. With a few clear exceptions, the system remained in the quiescent level during all of these other observations. To verify that the observed variations are intrinsic to the system rather than caused by observational artifacts, we show in Fig. \ref{fig:tess} the light curve of Gaia DR3 5460911351906171520, plotted in gray (bin of 2 minutes) and blue (bin of 90 minutes), for reference. This is a WD star observed by TESS on the same CCD as our target in each sector. This approach allowed us to exclude from the analysis data corresponding to spurious increases or decreases in flux at the beginning of four sequences - two in Sector 36 and two in Sector 62  - which are not included in the red and blue light curves. The discussion that follows is therefore attributed to the intrinsic variability of SPLUS J102139, given the absence of correlated variations in the reference star.

The most distinct event was detected at the end of the TESS coverage, in Sector 89, marked by green rectangles numbered in Fig.~\ref{fig:tess}, and is consistent with an outburst. The system initially exhibited a gradual decline in flux of $\sim$25\% over approximately 2.42 days (rectangle 1), reaching a normalized flux minimum of $\sim$0.75 and remaining in this low-brightness state for 1.70 days (rectangle 2). 
It was followed by a rapid brightening over 1.2 days, during which the normalized flux increased from $\sim$0.75 to $\sim$2.75, corresponding to an amplitude of nearly 2 magnitudes (rectangle 3). After reaching peak brightness, the system began to fade again (rectangle 4); however, the TESS coverage ended before the return to quiescence, leaving the decay phase of the outburst unrecorded. The observed $\sim 25$\% decrease in quiescent brightness and the remarkably short time scale of this event cannot be explained within the framework of thermal-viscous disc instabilities -- which only predicts a slow, steady increase in disc brightness over the long $\sim 180$\,d interoutburst timescale \citep[e.g.,][]{dubus2018} --, suggesting instead it may be part of the transient response of the accretion disk to a fast change in mass transfer rate driving the following outburst \citep[see, e.g.,][]{baptista2007}.

A second notable event is a flux drop - indicated by the green rectangle and letter A in Fig. \ref{fig:tess} - detected during the observation of Sector 36, between BJD 2285 and 2287. It corresponds to a decrease of approximately 25\% relative to the average flux estimated in 2.5 day intervals, one before and the other after the event. 
While this dip shares a comparable duration and flux deficit with the pre-outburst fading phase observed in Sector 89 (rectangle 1), its morphology is fundamentally different, presenting a gradual, triangular decline rather than a flat-bottomed profile.
As argued for the Sector 89 brightness fading event, a rapid flux drop of this magnitude during quiescence cannot be explained in the context of thermal-viscous disc instabilities.
Instead, it points to a brief reduction in the mass transfer rate from the donor star. It is worth noting that such a rapid and short-lived low state is highly unusual; typical low states driven by mass-transfer cessation in dwarf novae, such as the well-known case of HT Cas \citep{Wood1995}, tend to be considerably deeper and last significantly longer. The detection of this brief, triangular dip therefore highlights a peculiar, episodic mass-loss behavior from the companion star in SPLUS J102139.

The binned light curve also reveals low-amplitude, short-timescale variations consistent with flickering. In some cases, it additionally shows possible accumulations of flickering, highlighted by green rectangles and letters B and C in Fig. \ref{fig:tess}, in Sector 63. These enhanced variations on timescales of several orbits ($\sim$1--2 days, comparable to normal outbursts) may result from episodes of increased mass transfer or instabilities in the inner accretion disk, which lead to clusters of flickering flares occurring over these intervals.

\subsubsection{From OPD light curves}

\begin{figure}[]
\centering
\hspace*{-0.1cm}\includegraphics[width=0.5\textwidth]{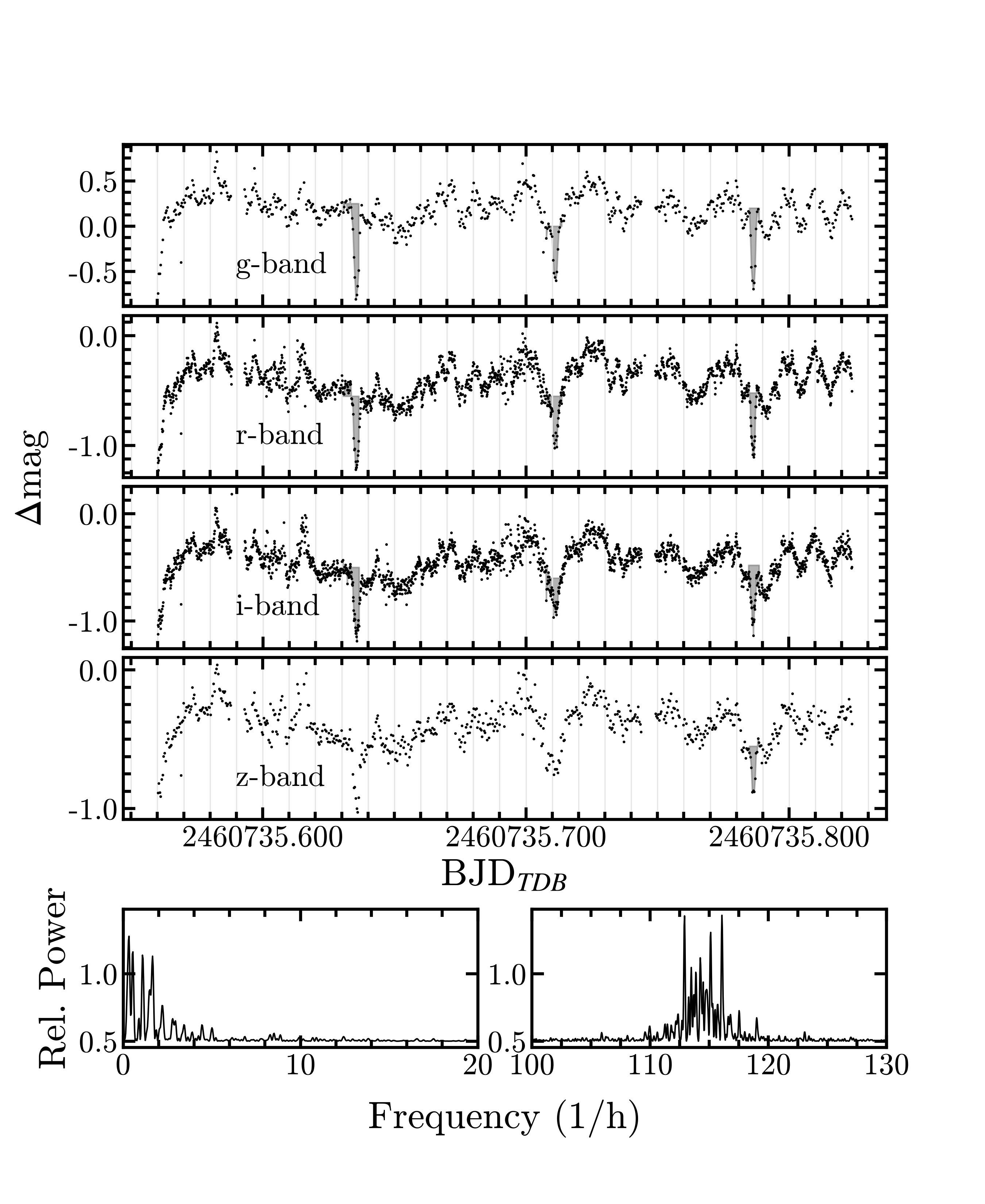}
\caption{SPARC4 $griz$ light curves and Lomb-Scargle Periodogram of the $r$ band SPARC4 light curve in two frequency ranges.}
\label{fig:sparc4}
\end{figure}

The OPD light curves exhibit irregular short-timescale variations, commonly referred to as flickering. These variations are observed across all orbital phases and are typical of accretion-driven systems. To investigate the temporal characteristics of the flickering and gain insight into its physical origin, we computed the Lomb–Scargle periodogram of the residual light curve, obtained after subtracting a low-order trend via spline fitting. 

The Lomb–Scargle periodogram (Fig. \ref{fig:sparc4}) reveals a modest peak at low frequency (\(f_{\rm orb} \approx 0.55\,\mathrm{h}^{-1}\)), associated with the orbital modulation and the occurrence of eclipses. However, this peak is not significantly stronger than other low-frequency structures in the spectrum.

In the intermediate frequency range (\(6\)–\(12\,\mathrm{cycles\,h}^{-1}\)), several low-amplitude peaks are detected. These features are simply characteristic of the red noise typically associated with stochastic flickering in cataclysmic variables.

At higher frequencies, a cluster of signals around \(f \approx 113\,\mathrm{h}^{-1}\) (corresponding to a timescale of \(\sim 32\,\mathrm{s}\), lower right panel of Fig.~\ref{fig:sparc4}) clearly indicates the presence of short-period QPOs (see \citealt{warner2004} for a characterization of rapid oscillations in CVs).

Overall, the power spectrum does not reveal coherent periodic signals, but supports the stochastic origin typical for the flickering, possibly with contributions from multiple regions in the accretion structure.

To explore the wavelength dependence of the flickering, we measured the standard deviation of the residuals in each of the four OPD bands as a proxy for the flickering amplitude. The measured dispersions are nearly identical ($\sim$0.0316 mag), indicating that the flickering amplitude does not vary significantly with wavelength. This result implies that the rapid brightness fluctuations are produced by a broadband emission process, likely originating from a thermally dominated, optically thick region of the accretion disk.

\subsection{Orbital Modulation and Eclipses}
\label{sect:orbmodul}

We analyzed the TESS light curves of SPLUS J102139 to assess periodic variability in the system. The goal was to identify signatures consistent with orbital modulation, such as flux reductions caused by eclipses and WD spin modulation. For this purpose, the Lomb-Scargle periodogram was applied to the light curve of each sector, separately (Fig.~\ref{fig:tess}).

The power spectra show clear signals of a periodic oscillation, including its fundamental period ($P_{\Omega}$) and its first five harmonics ($P_{\Omega/2}$, $P_{\Omega/3}$, $P_{\Omega/4}$, $P_{\Omega/5}$, and $P_{\Omega/6}$), as presented in the middle panel of Fig.~\ref{fig:tess}. The $P_{\Omega}$ values derived from the observations in sectors 36, 62, 63, and 89 by fitting the peaks with a Gaussian are, respectively: $P_{\Omega,\rm{S36}} = 1.8101 \pm 0.0019$ h, $P_{\Omega,\rm{S62}} = 1.8100 \pm 0.0019$ h, $P_{\Omega,\rm{S63}} = 1.8110 \pm 0.0019$ h, and $P_{\Omega,\rm{S89}} = 1.8100 \pm 0.0017$ h. These values are mutually consistent within their uncertainties and correspond to the orbital period of SPLUS J102139. By computing the weighted mean of the four measurements, where the weights are based on the inverse squares of their uncertainties, we estimate a final orbital period of \(P_\Omega = {1.810 \pm 0.001}\,\mathrm{h}\). The folded light curves at $P_{\Omega}$, binned into phase intervals of width 0.0125 to improve the signal-to-noise ratio and enhance the visibility of the periodic modulation, show that the modulation comes from recurrent eclipses (bottom panel in Fig. \ref{fig:tess}). In these eclipses, the flux decreases by approximately 17 to 24\%. Due to the potential dilution of the light curve by background contamination in the large TESS aperture, the residual flux level observed at mid-eclipse should be treated as an upper limit to the actual minimum flux of the cataclysmic variable.
The presence of a noticeable substructure between eclipses, interpreted as being at least partially due to ellipsoidal modulation of the secondary, accounts for the first harmonic appearing stronger than the fundamental in the periodogram.  
The periodogram shows no evidence of a signal corresponding to WD's spin period. The eclipses are also clearly visible in the OPD light curves (Fig. \ref{fig:sparc4}). The time interval between consecutive eclipse minima, measured from successive ingress times, is 1.81 hours, in agreement with the orbital period derived independently from TESS photometry. 

We used the TESS observations to refine the orbital period and to obtain a precise
ephemeris for the mid-eclipse times using average, representative eclipse timings
from combined large number of eclipse cycles \citep[e.g.,][]{baptista2003}. First,
we adopted a test ephemeris assuming $P_\Omega$ as the orbital period, phase-folded
the data of the four TESS sectors with this ephemeris, and combined all orbits
of each sector into a median orbital light curve. We then computed an
observed-minus-calculated (O$-$C) value for each median light curve by measuring
the phase offset between the observed mid-eclipse and the phase zero of the
test ephemeris. We transformed each measured (O$-$C) value into an eclipse timing
by selecting a representative eclipse cycle $E$ for each TESS sector and adding
the observed (O$-$C) value to the corresponding mid-eclipse timing predicted by
the test ephemeris. The resulting four representative mid-eclipse timings are
listed in Table\,\ref{timings}.

\begin{table}
  \caption{Mid-eclipse timings of SPLUS J102139.}
  \begin{tabular}{ccc}
  \hline \hline
  E & BJD & (O$-$C)
  \footnote{with respect to the ephemeris of Eq.(\ref{efemeride}).}\\
  (cycle) & (2.459.000+) & (cycle) \\ [2pt]
   \hline
   156 &  293.94795 (8) & $-0.00077$ \\
  9533 & 1001.14917 (8) & $+0.00105$ \\
  9890 & 1028.07360 (8) & $+0.00049$ \\
 19208 & 1730.37238 (8) & $-0.00077$ \\
  \hline
  \end{tabular}
\label{timings}
\end{table}
The best-fit linear ephemeris to these timings is,
\begin{equation}
    T_\mathrm{mid}= \mathrm{BJD}\, 2\,459\,282.182\,69 (7)
    + 0.075\,418\,693 (6) \times E \, ,
    \label{efemeride}
\end{equation}
where $E$ is the cycle number.

\subsection{Binary parameters}

\begin{figure}
    \centering
        \includegraphics[width=\linewidth]{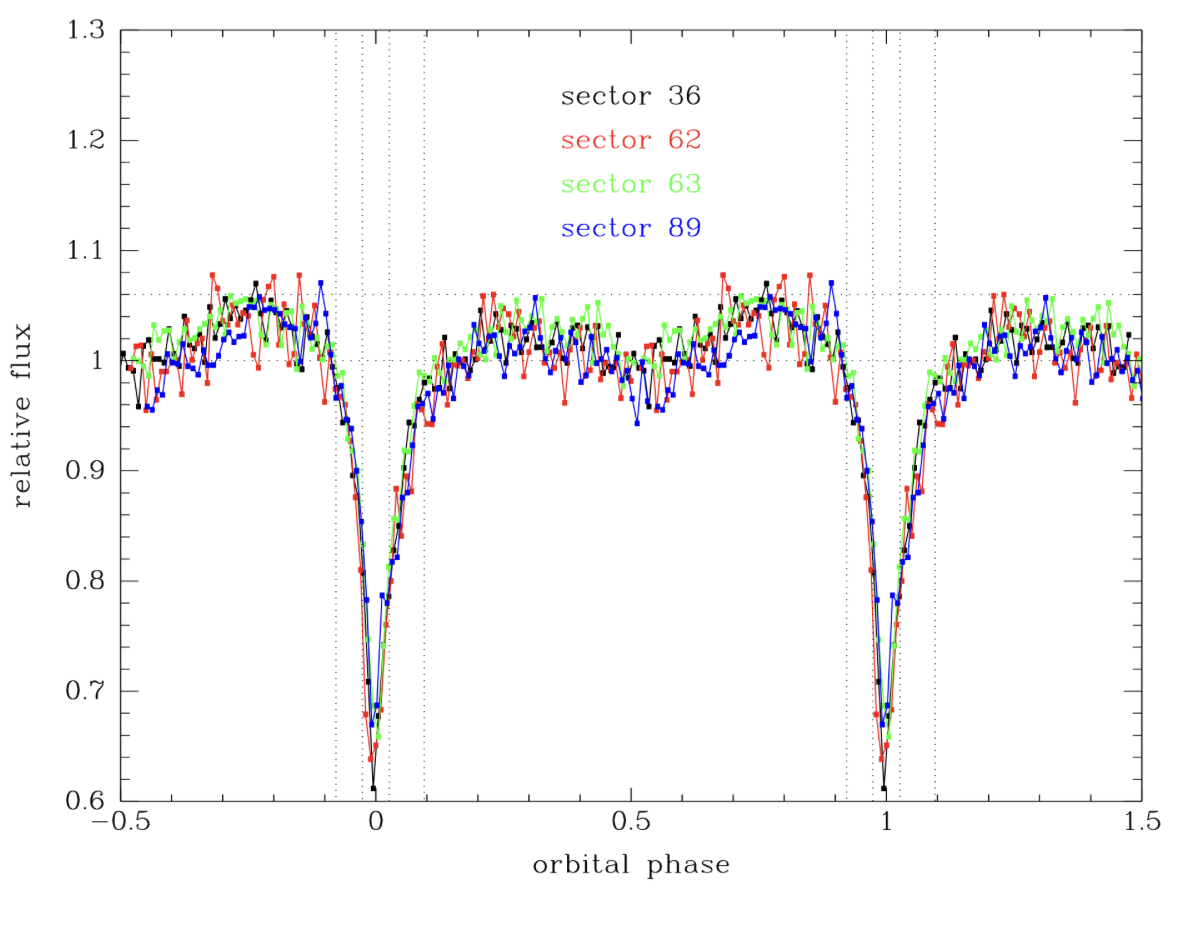}
    \caption{ 
    Phase-folded median light curve of SPLUS J102139 combining data from TESS 
    Sectors 36, 62, 63, and 89 (points), after subtraction of an additive
    component to the flux. Vertical dotted lines mark the total width of the
    eclipse, $\Delta\phi_E= 0.173$ in phase, and the width of the eclipse of
    the disk center, $\Delta\phi= 0.029$ in phase.}
    \label{fig:eclipse_mapping_fit}
\end{figure}

A photometric analysis of SPLUS J102139 was conducted using the TESS data.
Initial processing involved the normalization of light curves to the median SAP flux of sector 63, the one with the lower flux value. Comparison across all sectors reveals consistent orbital modulations, characterized by a prominent hump and ellipsoidal variations from the secondary star. The morphology of the eclipse -- specifically its width, depth, and shape -- remains stable across sectors, suggesting that observed flux differences are attributable to an additive component. By subtracting this component, the sectors were co-added to produce a high signal-to-noise (S/N) median combined light curve (see Fig.\,\ref{fig:eclipse_mapping_fit}).

The observed eclipse shape is smooth, shallow, and asymmetric, notably lacking the sharp transitions typically associated with the occultation of compact sources such as the WD or a bright spot at disk rim. The absence of these sharp eclipse features precludes a purely photometric determination of the binary geometry. Consequently, the physical parameters were initially constrained using the semi-empirical evolutionary sequences of \citet{knigge2011}. 

From the orbital period, the secondary mass was estimated as $M_2 = 0.127 \, M_\odot$. Assuming a primary mass range of $M_1 = 0.65-0.85 \, M_\odot$ \citep[consistent with a $2\sigma$ uncertainty relative to the median mass of WDs in cataclysmic variables,][]{knigge2011}, we estimated an initial mass ratio of $q = M_2/M_1= 0.17 \pm 0.04$.

We used the maximum-entropy eclipse mapping method \citep{Horne1985,baptista2016}
to infer the orbital inclination $i$ and to refine the mass ratio
estimate with the aid of the ''entropy landscape'' procedure
\citep[e.g.,][]{baptista2026}. The eclipse map of maximum entropy is the one
with the most axi-symmetric disk surface brightness distribution consistent
with the observed eclipse shape. A pair of ($i,q$) values define an eclipse
geometry. For a given $q$ value, the surface brightness distribution of an
eclipse mapping reconstruction is skewed towards (away from) the inner
Lagrangian point L1 if the inclination is underestimated (overestimated).
Thus, an incorrect inclination increases the amount of structure and the
deviations from axi-symmetry of the eclipse map, which is flagged with a 
lower entropy value. Hence, the best choice for the eclipse geometry can be
inferred by computing eclipse maps for a range of ($i,q$) values and finding
the pair of values that yields the eclipse map of highest entropy. 

\begin{figure}
    \centering
    \hspace*{-0.5cm}
    \includegraphics[width=0.65\textwidth]{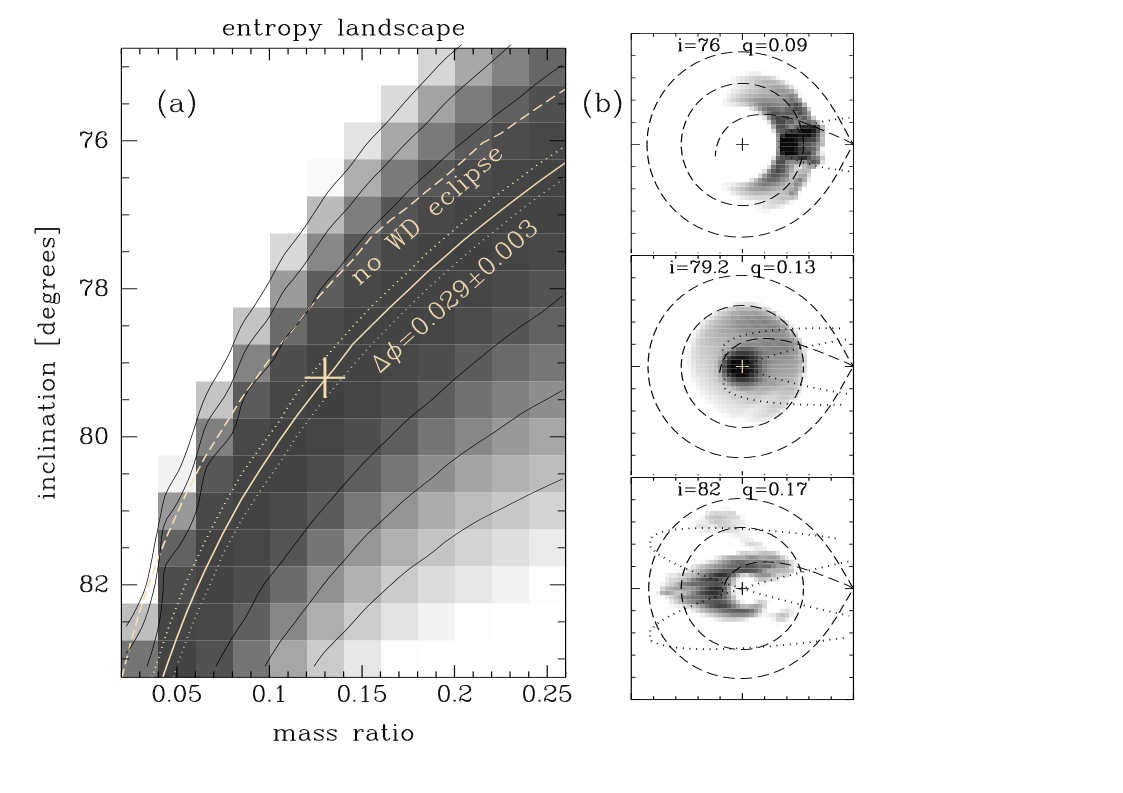}
    \caption{(a) Entropy landscape in the $(i,q)$ binary parameter plane. The entropy of the eclipse maps is shown as different shades of gray; darker regions have higher entropy. Contour lines indicate the shape of the entropy landscape and a white cross marks the location of highest entropy. White curves show lines of constant $\Delta\phi$ values and the line above which there is no eclipse of the WD at disk center (dashed curve). (b) Eclipse maps in a logarithmic grayscale for pairs of specific $(i,q)$ values; darker regions are brighter. A cross marks the center of the disk; dashed lines show a disk of radius $R_d= 0.55\,R_{L1}$ as well as the primary Roche lobe and the ballistic stream trajectory for the corresponding $q$ value. The secondary star is to the right of each panel and the stars (and the
    disk gas) rotate in the counter-clockwise sense. Dotted lines illustrate the projected shadow of the secondary star onto the orbital plane at the ingress/egress phases of the disk center in each case; because there is no eclipse of disk center in the upper panel, the shadow at phase zero is shown.}
    \label{fig:entropy_landscape}
\end{figure}

For the entropy landscape procedure, we used the high S/N median combined light
curve of all sectors. Because the employed version of the eclipse mapping method
does not take into account out-of-eclipse brightness changes, these were removed
by dividing the median light curve by the best-fit spline function to the
out-of-eclipse portion of the light curve, and scaling the result to the
average out-of-eclipse flux level. The results of the entropy landscape are
shown in Fig.\,\ref{fig:entropy_landscape}. The maximum entropy solution yields a mass ratio of $q = 0.13 \pm 0.04$ and
an orbital inclination of $i = 79.2^\circ \pm 1.5^\circ$. This corresponds to
an eclipse duration of the disk center of $\Delta\phi = 0.029 \pm 0.003$,
implying a grazing eclipse of both the WD and the central disk regions.
The $q$ value inferred from the entropy landscape is consistent with that
estimated from evolutionary considerations at the 1$\sigma$ limit.
This solution also identifies a non-eclipsed flux component accounting
for approximately 30\% of the total median flux.
The right hand panels show eclipse maps obtained for the correct eclipse
geometry (middle panel) as well as for two inconsistent geometries that
move the brightest part of the surface distribution towards (upper panel)
and away (lower panel) from L1. The eclipse maps with inconsistent eclipse
geometry are plagued with criss-crossed arcs, indicating that noise in the
light curve is transposed to the eclipse map as the entropy and the data
constraints are ``fighting'' each other.
The eclipse map of highest entropy shows an extended central source 
about 4-5 times larger than the expected size of the WD. 
The accretion disk shows enhanced emission at the quadrant containing the gas stream (responsible for the asymmetric eclipse shape) but no clear evidence of a compact bright spot at disk rim.

From the median combined light curve we measure a total eclipse duration
of $\Delta\phi_E= 0.173 \pm 0.002$ in phase, corresponding to an outer
disk radius of $R_d/a = 0.387 \pm 0.012$, or $R_d/R_{L1} = 0.55 \pm 0.02$.
The WD mass was inferred from the estimated $M_2$ and $q$ values; the orbital
separation $a$ follows from Kepler's third law. The distance from the WD to the
L1 point is obtained from the expression of \citet{Plavec64}, the WD radius is
derived from the mass-radius relation of \citet{Nauenberg72}, and $R_2$ is taken
as the volume equivalent Roche lobe radius of the secondary star \citep{Eggleton83}.
A standard error propagation routine using Monte Carlo simulation of $10^5$ trials
with the input parameters $q$, $M_2$ and $\Delta\phi_E$ is used to estimate the
uncertainties of the inferred parameters. The results are summarized in
Table\,\ref{tab:sys_params}.

\begin{table}[ht!]
\centering
\caption{Binary parameters of SPLUS J102139}
\label{tab:sys_params}
\begin{tabular}{lc}
\hline\hline
Parameter & Value \\
\hline
$q$ & $0.130 \pm 0.040$ \\
$i$ &  $79.2^\circ \pm 1.5^\circ$ \\
$M_1$ ($M_\odot$) & $0.968 \pm 0.068$ \\
$R_1$ ($R_\odot$) & $0.0082 \pm 0.0008$ \\
$M_2$ ($M_\odot$) & $0.127 \pm 0.043$ \\
$R_2$ ($R_\odot$) & $0.172 \pm 0.021$ \\
$R_d/a$ & $0.387 \pm 0.012$ \\
$a$ ($R_\odot$) & $0.773 \pm 0.024$ \\
$K_1$ (km/s) & $58.3 \pm 17.0$ \\
$K_2$ (km/s) & $451.6 \pm 12.2$ \\
\hline
\hline
\end{tabular}
\end{table}

\subsection{Characterization Based on Swift/UVOT Observations}

\begin{table*}[ht!]
    \scalefont{0.8}
    \centering
    \begin{tabular}{cccccccccc}
         \hline
         \hline
         ObsID & Date & XRT Exp. & UVOT Exp. & Filter & Extname & Magnitude & Flux & Flux unabs. & Flux unabs.\\
         && (s) & (s)& & & (mag) & & x-ray & map\\

         \hline
         \multirow{1}{*}{00015951001} &  2023 April 15 & 180.486  & 180.035  & UVW2 & w2703250400I & 15.85$_{-1.09}^{+1.09}$ & $<48.85$ & $<755.65$ & $<63.27$\\
         \hline
         \multirow{4}{*}{00015951002} & \multirow{4}{*}{2023 April 21} & \multirow{4}{*}{3297.100}  & \multirow{4}{*}{3304.734}  & \multirow{4}{*}{UVW1} & w1703729347I & 16.56$_{-0.06}^{+0.06}$ & 9.48$_{-0.55}^{+0.55}$ & 86.72$_{-5.03}^{+5.03}$ & 11.68$_{-0.68}^{+0.68}$\\
         & & & & & w1703735037I & 16.49$_{-0.05}^{+0.05}$ & 10.10$_{-0.50}^{+0.50}$ & 92.39$_{-4.57}^{+4.57}$ & 12.45$_{-0.62}^{+0.62}$ \\
         & & & & & w1703740749I & 16.46$_{-0.05}^{+0.05}$ & 10.30$_{-0.45}^{+0.45}$ & 94.22$_{-4.12}^{+4.12}$ & 12.70$_{-0.55}^{+0.55}$\\
         & & & & & w1703758224I & 16.56$_{-0.08}^{+0.08}$ & 9.48$_{-0.70}^{+0.70}$ & 86.72$_{-6.40}^{+6.40}$ & 11.68$_{-0.86}^{+0.86}$ \\
         \hline
         \multirow{2}{*}{00015951003} & \multirow{2}{*}{2023 April 26} & \multirow{2}{*}{739.653}  & \multirow{2}{*}{1370.432}  & \multirow{2}{*}{U} & uu704182392I & 18.66$_{-1.62}^{+1.62}$ & 1.22$_{-1.81}^{+1.81}$ & 6.32$_{-9.38}^{+9.38}$ & 1.43$_{-2.11}^{+2.11}$ \\
         & & & & & uu704239760I & 18.42$_{-1.40}^{+1.40}$ & 1.51$_{-1.96}^{+1.96}$ & 7.83$_{-10.16}^{+10.16}$ & 1.76$_{-2.29}^{+2.29}$\\
         \hline

    \end{tabular}
    \caption{Summary of Swift observations, the UVOT magnitudes and fluxes, and the unabsorbed X-ray fluxes (in units of 10$^{-16}$ erg\,s$^{-1}$cm$^{-2}$\AA$^{-1}$).}
    \label{tab:swiftdata}
\end{table*}

We explore the ultraviolet/optical photometry of SPLUS J102139 obtained with Swift/UVOT, as summarized in Table~\ref{tab:swiftdata}. The observations include measurements in three filters: UVW2, UVW1, and U, acquired on different dates. 

The UVW1 data include a set of four exposures obtained during a single observing sequence. The resulting magnitudes are mutually consistent within uncertainties, indicating no detectable variability over the short interval spanned by the observations (3.3 ks).  Observations in UVW2 and U were each limited to one and two exposures, respectively, and are affected by significantly larger photometric uncertainties. Furthermore, cross-filter comparisons are not straightforward due to the differing effective wavelengths and extinction responses of each filter.

Using this E(B–V)=0.034 mag (see Sect. \ref{sct:gaia}), the extinction curve by \cite{cardelli} was employed to determine the wavelength-dependent extinction. Specific coefficients for each UVOT filter, provided by \cite{Roming_2009}, were applied to calculate the extinction $A_{\lambda}$. The corrected fluxes were then determined as $F_{\text{corr}} (\lambda) = F_{\text{obs}} (\lambda) \times 10^{0.4 \times A^{\lambda}}$, providing a more accurate estimate of the intrinsic emission of the source (see Table \ref{tab:swiftdata}).

To quantify the relative contribution of ultraviolet (UV) and X-ray emission, we computed the integrated UV flux in the UVW1 band by multiplying the extinction-corrected flux densities by the effective width of the filter (693\,\AA; \citealt{Poole2008}). This resulted in UV fluxes of \(8.10 \times 10^{-13}\), \(8.63 \times 10^{-13}\), and \(8.80 \times 10^{-13}\,\mathrm{erg\,s^{-1}\,cm^{-2}}\). When compared to the unabsorbed X-ray flux of \(2.86 \times 10^{-12}\,\mathrm{erg\,s^{-1}\,cm^{-2}}\) (see Section~\ref{sect:xrt_results}), the corresponding X-ray-to-UV flux ratios are approximately 3.5, 3.3, and 3.2, respectively. These values indicate that the high-energy emission in this system is significantly dominated by the X-ray component.

\subsection{X-ray Spectroscopy Analysis}
\label{sect:xrt_results}

\begin{figure}[t!]
\centering
\includegraphics[width=8.5cm,scale=0.1]{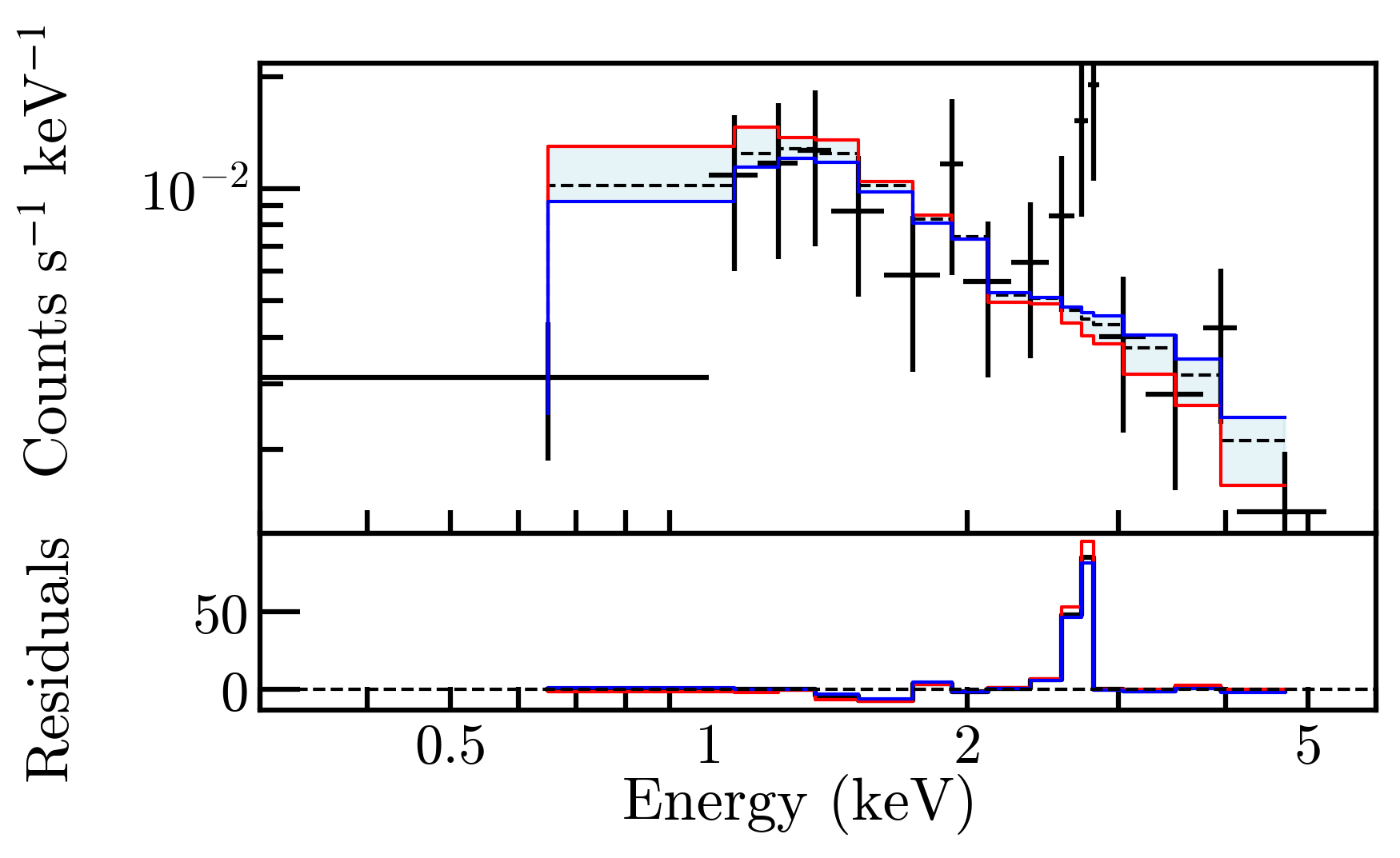}
\caption{Swift X-ray spectrum of SPLUS J102139. The black dashed line represents the best-fit model (\texttt{tbabs*apec}). The red and blue solid lines show alternative fits with fixed plasma temperatures of $kT = 5$ keV and $kT = 20$ keV, respectively. The shaded light blue area marks the envelope between these two models. Residuals are shown in the bottom panel.}
\label{fig:xspec}
\end{figure}

The source SPLUS J102139 was detected in X-rays with Swift/XRT, with a total net count rate of $(3.23 \pm 0.35) \times 10^{-2}$ counts s$^{-1}$ in the 0.3 - 10 keV energy range. Following the low-count X-ray spectra, we employed C-statistics  \citep[Cash statistics;][]{Cash1979}) for the spectral fitting, as it provides a more reliable approach in the Poisson regime without requiring Gaussian errors. The C-statistic was used throughout the modeling to obtain best-fit parameters and their confidence intervals. For comparison and model goodness assessment, we also report the reduced $\chi^2$ and degrees of freedom ($\chi^2_\nu$/d.o.f.), which offer a standard measure of the overall fit quality, even though $\chi^2$ was not used as the minimization statistic.

The spectral model applied combines two fundamental components: x-ray absorption and thermal emission, represented by the {\fontfamily{qcr}\selectfont
tbabs} and {\fontfamily{qcr}\selectfont
apec} models, respectively. The {\fontfamily{qcr}\selectfont
tbabs} component accounts for absorption by neutral interstellar material along the line of sight, quantified by the equivalent in hydrogen column density ($N_H$). The {\fontfamily{qcr}\selectfont
apec} model describes emission from a hot, optically thin plasma in collisional ionization equilibrium, with main parameters including the plasma temperature ($kT$), metallicity, redshift, and normalization.

We adopted the abundance table from \cite{aspl}. Unconstrained, the abundance was fixed to the solar values. Although galactic sources span a range of metallicities and the Sun is not necessarily representative of all populations, adopting solar abundances is a practical choice given the limited spectral signal, which does not support fitting this parameter. We also froze the redshift at 0, as appropriate for a non-extragalactic source. Fixing these parameters reduces model complexity and stabilizes the estimation of the plasma temperature and unabsorbed flux.

Applying the model {\fontfamily{qcr}\selectfont
(tbabs*apec)}, we find a hydrogen column density $N_{H} = 0.67_{-0.22}^{+0.40} \times 10^{22} \,cm^{-2}$. However, the plasma temperature was unconstrained. The confidence interval yields unphysical negative lower bounds, indicating that the data do not provide sufficient sensitivity to determine 
$kT$. This model provides a reduced chi-squared value of $\chi_{v}^{2}/d.o.f = 0.99/14$. Figure \ref{fig:xspec} shows the spectrum along with the best-fit model and residuals, which represent the deviations between the observed data and the prediction of the fitted model. A localized positive residual is visible at energies close to 3 keV, indicating an excess of counts relative to the single-temperature plasma model. Given the limited number of counts and the moderate spectral resolution of Swift/XRT, however, the statistical significance and physical origin of this feature cannot be reliably established. It may represent a statistical fluctuation or unresolved line emission not adequately reproduced by the adopted model. Higher signal-to-noise X-ray spectroscopy is required to determine whether this excess is intrinsic to the source. Neglecting the small interstellar absorption (\( N_{\mathrm{H}} = 0 \)), the unabsorbed flux in the 0.3-10~keV energy range is 
\( 2.86 \times 10^{-12}\,\mathrm{erg\,cm^{-2}\,s^{-1}} \). Adopting a distance of 499~pc (see Section \ref{sct:gaia}), this corresponds to an X-ray luminosity of 
\( 8.5 \times 10^{31}\,\mathrm{erg\,s^{-1}} \).

To further explore the spectral shape given the unconstrained plasma temperature, we performed two additional fits using the same {\fontfamily{qcr}\selectfont
(tbabs*apec)} model. In both cases, we fixed the hydrogen column density as the best-fit obtained previously, and set the plasma temperature to fixed values $kT = 5$ keV and $kT = 20$ keV, respectively. These temperatures were chosen to span a plausible range, considering the lack of constraint in the original fit. The resulting models define a region that helps visualize the possible spectral shapes consistent with the data in the 0.3 - 10 keV. These fixed temperature models are shown in Fig. \ref{fig:xspec}, in which the red and blue lines correspond to $kT = 5$ keV and $kT = 20$ keV, respectively. The light blue shaded area highlights the region between these two extremes.

\section{Discussion and Final Remarks}

We have conducted a detailed optical and UV timing, and optical and X-ray spectral analysis of the object SPLUS J102139, revealing characteristics consistent with its classification as a short-period eclipsing dwarf nova. This conclusion is supported by several key observations: the short orbital period of 1.81 hours, the outburst behavior, the absence of a detectable WD spin period, a X-ray luminosity of $8.5 \times 10^{31} \,$ \ergs \,in the 0.3–10 keV range, and optical spectral signatures.

The optical spectrum of SPLUS J102139 is dominated by broad, double-peaked emission lines, a hallmark of an edge-on accretion disk system. These features arise from the Doppler broadening of emission originating in the disk's inner and outer regions, with peaks corresponding to blue- and red-shifted material moving at different velocities \citep{greenstein}. The identification of He\,II 4686 \AA\ and H$\beta$ emission lines in the spectrum provides additional diagnostics for the system's accretion environment. He II emission line requires high energy photons capable of ionizing He II ions ($>$54 eV).  Such photons are routinely produced by magnetic CVs although they can also be produced in the boundary layer of non-magnetic CVs \citep{silber}. This is consistent with the absence of strong magnetic features, ruling out the possibility of a polar classification.  
While the presence of double-peaked emission supports the existence of a Keplerian accretion disk, a feature characteristic of non-magnetic dwarf novae \citep{patterson}, we cannot completely discard the possibility that the system is a low-luminosity intermediate polar (LLIP).  Indeed, some LLIPs exhibit outbursts and superoutbursts that closely resemble those of non-magnetic dwarf novae, along with spin modulations that are difficult to detect due to their short periods or low amplitudes. Notable examples of such hybrid behavior include CC Scl \citep{woudt2012, FeII_1} and CTCV J2056-3014 \citep{lopes, hameury2022}.

The presence of Fe\,II 5169 \AA\ emission, belonging to multiplet 42, sheds light on the physical conditions within the accretion environment of SPLUS J102139. This feature is believed to originate from a UV-driven fluorescence mechanism, in which high-energy photons illuminate a dense, vertically extended gas layer above the disk, triggering Fe\,II transitions that emit in the optical range \citep{mason2005}. Previous studies indicates that Fe\,II features tend to emerge in post-outburst stages and are typically absent during quiescence \citep{williams1983, papadaki2008}. In this context, the detection of Fe\,II 5169 \AA\ in SPLUS J102139 could signify residual heating or lingering accretion activity following a recent outburst, supporting its classification as a non-magnetic dwarf nova observed in a mildly active state.

Each individual spectrum integrates light over about 9.2\% of the orbital cycle. As a consequence, the observed spectral features represent a time-averaged view of the system during different orbital phases. Since the system continues to rotate during the exposure, the resulting spectrum may combine contributions from distinct dynamical configurations, such as varying viewing angles of the accretion disk, the bright spot, or asymmetric emission regions. This integration over a non-negligible orbital phase implies that some velocity smearing or line profile blending may occur, particularly in systems with rapid orbital motion. Furthermore, features such as double-peaked emission lines may be partially averaged out, potentially masking transient or anisotropic structures. Therefore, while the spectra are valuable for confirming the accreting nature and emission-line characteristics of the system, they must be interpreted as phase-blended snapshots, and not as instantaneous views of a fixed orbital geometry. 

The ASAS-SN survey has provided a valuable long-term dataset, capturing 16 brightness increase events over a decade, with an average occurrence interval of approximately 180 days. However, seasonal gaps in survey coverage indicate that the true frequency of such events may be underestimated. This highlights the challenges inherent in using ground-based surveys with periodic observational constraints, particularly for systems with short-lived or infrequent outbursts. The March 2019 event (Event 6) is particularly significant, as the system's brightness peaked at magnitude 14.9 and remained above magnitude 17.0 for over 11 days. This extended duration aligns with the characteristic behavior of superoutbursts driven by the thermal-tidal instability in the accretion disk \citep{Osaki1996}. This behavior provides a critical connection to the system's orbital period of 1.81 hours. Dwarf novae with orbital periods shorter than 2 hours are subject to tidal instabilities because their expanding disks can reach a resonant state with the companion star. These short-period systems are generally divided into two categories: normal SU UMa systems, which exhibit both normal outbursts and superoutbursts, or WZ Sge-type systems, a special subset that exhibits only superoutbursts. While the detection of both numerous short, normal outbursts and this prominent superoutburst phenomenologically classifies SPLUS J102139 as a possible SU UMa-type dwarf nova -- it is important to note that some LLIPs can mimic this outburst behavior (as further discussed above).

Such superoutbursts are believed to result from increased disk viscosity and tidal interactions, leading to an enhanced transfer of angular momentum and mass within the disk. In contrast, Events 12 and 13 (recorded in March and December 2022, respectively)  exhibit intermediate timescales that distinguish them from both normal outbursts and the classic superoutburst. Although the limited sampling of the ASAS-SN data does not permit a detailed resolution of their light curve morphology, these extended durations might reflect variations in accretion dynamics, potentially driven by changes in the mass transfer rate from the companion star or the impact of disk instabilities that do not fully align with standard models \citep{Lasota2001}. 

The TESS and OPD light curves reveal a short orbital period of 1.81 hours, consistent with expectations for low-luminosity CVs \citep{knigge2011, Pretorius}.  Unlike typical high-inclined systems, the photometry of SPLUS J102139 shows smooth, shallow, and asymmetric eclipses, notably lacking the sharp transitions associated with the occultation of a compact WD or a well-defined bright spot. This grazing eclipse geometry necessitated the use of an entropy landscape procedure to precisely constrain the system parameters. The maximum entropy solution revealed an extreme mass ratio of $q = 0.13 \pm 0.04$ and an orbital inclination of $i = 79.2^\circ \pm 1.5^\circ$. This robust geometric constraint allowed for accurate mass determinations for the primary ($M_1 = 0.968 \pm 0.068 \, M_\odot$) and secondary ($M_2 = 0.127 \pm 0.043 \, M_\odot$) components. Furthermore, the eclipse mapping reconstructs a well-behaved central source and an asymmetric accretion disk that extends to $R_d = 0.55 \pm 0.02 \, R_{L1}$. Although there is a slight brightness enhancement in the bright spot quadrant, no compact source is clearly resolved at the disk edge. However, it is important to note that TESS operates in a broad red-to-near-infrared passband. In this wavelength regime, the contrast of the typically hot, blue bright spot against the cooler outer disk is substantially reduced compared to visual observations. The lack of a sharply defined compact source in this passband, combined with the grazing nature of the central eclipse, elegantly explains the absence of rapid flux ingress or egress in the TESS light curve. Finally, the absence of any coherent spin signal in the TESS periodograms further supports the non‑magnetic dwarf nova classification \citep{Horne1985}.

The Swift/UVOT observations show a stable UV flux during the April 2023 exposures, with extinction-corrected values reaching up to $\sim (8.10-8.80) \times 10^{-13}$ erg\,s$^{-1}$\,cm$^{-2}$ in UVW1. These values are consistent with a quiescent accretion state, in which the boundary layer emits in X-rays while the disk contributes to the UV \citep{verbunt1997}. The X-ray to UV flux ratio ($\sim$3.2–3.5) is consistent with the behavior of dwarf novae in quiescence, where the boundary layer remains optically thin and radiates efficiently in the X-ray regime, while the accretion disk contributes modestly to the UV flux. This contrasts with nova-like systems or dwarf novae during outburst, in which the optically thick disk dominates the radiative output and the X-ray emission is suppressed \citep{patterson1985,verbunt1997}. Therefore, the observed ratio supports the interpretation that the system was observed in a low accretion state.
  
The spectral fit in X-rays is statistically robust, resulting in $\chi_{v}^{2}$ $/\mathrm{d.o.f} = 0.99/14$. The X-ray luminosity is estimated in $8.5 \times 10^{31}$ \ergs\ in the 0.3–10 keV range, near to the upper limit of dwarf novae, comparable to SS Cyg, which defines the upper boundary of X-ray luminosity for this class following \citet{Byckling2010}. The luminosity suggests the system is in quiescence, with accretion occurring at a reduced rate and emission arising from the boundary layer between the accretion disk and the WD \citep{Narayan1994}. While both LLIPs and dwarf novae typically exhibit low luminosities, the absence of a detectable spin period in TESS data makes the classification of the system as a dwarf nova more likely.

The identification of this object as an eclipsing dwarf nova is intriguing and warrants further investigation. Eclipsing dwarf novae are much rarer than their non-eclipsing counterparts. Although \citet{Kozhevnikov2021} identified 108 eclipsing dwarf novae with measured eclipse depths in the VSX database and the literature, only a smaller fraction are sufficiently bright in quiescence for detailed follow-up observations. The catalog compiled by \citet{Mukai2009}, which focuses on relatively bright eclipsing dwarf novae suitable for X-ray studies, contains about 30 systems, of which only around a dozen have short orbital periods. Such systems are particularly valuable for detailed characterization, allowing precise measurements of system parameters such as inclination, mass ratio, and accretion disk geometry, which are critical for understanding binary evolution \citep{warner}.
 
Although the CV nature of SPLUS J102139 was already known by ASAS-SN survey, our analysis provides a comprehensive multi-wavelength characterization of the system, including new optical spectroscopy, time-resolved photometry, and X-ray/UV data. The independent discovery and characterization of SPLUS J102139 exemplify the power of photometric surveys like S-PLUS to uncover CVs that may be overlooked by traditional selection techniques.
This approach not only expands the known CV population, particularly at lower luminosities, but also contributes to refining the classification and understanding of diverse CV subtypes.

\begin{acknowledgments}
A.S.A. was supported by {\it Coordenação de Aperfeiçoamento de Pessoal de Nível
Superior} (CAPES; Finance Code 001: 88887.705415/2022-00
and 88887.992353/2024-00) and {\it Fundação Carlos Chagas
Filho de Amparo à Pesquisa do Estado do Rio de Janeiro}
(FAPERJ; E-26/203.109/2023-29438).
R.L.O. was partially supported by {\it Conselho Nacional de Desenvolvimento Científico e Tecnológico} (CNPq; PQ-315632/2023-2 and 445047/2024-0), {\it Fundação de Amparo à Pesquisa do Estado de São Paulo} (FAPESP; 2025/01204-6), and acknowledges financial support from NASA and the XRISM Project Science Office during his stay at NASA’s Goddard Space Flight Center {\bf in 2026}. 
R.E. was supported by CAPES (Finance Code 001: 88887.914708/2023-00).
Partially based on observations carried out at the Observatório do Pico dos Dias (OPD), operated by the Laboratório Nacional de Astrofísica (LNA/MCTI), under OPD observing program 2025A-P-022 (PI: R. Eleutério).
\end{acknowledgments}

\bibliography{ref}
  
\end{document}